\documentclass{jfm}
\usepackage{graphicx}
\usepackage{epstopdf, epsfig}
\usepackage{amsmath}
\usepackage{natbib}
\defcitealias{barthelemy2018on-a-unified-br}{B18}

\shorttitle{On the energetics of breaking inception and onset in surface gravity waves}
\shortauthor{D. G. Boettger et al}

\title{On the energetics of breaking inception and onset in surface gravity waves}

\author{Daniel G. Boettger\aff{1}
  \corresp{\email{d.boettger@student.unsw.edu.au}},
  Michael L. Banner\aff{1},
  Xavier Barth\'el\'emy\aff{1},
  Shane R.  Keating\aff{1},
  \and Russel P. Morison\aff{1}}

\affiliation{\aff{1} School of Mathematics and Statistics, University of New South Wales,
Sydney, Australia}

\begin{document}

\maketitle

\begin{abstract}
Accurate prediction of the onset and strength of breaking surface gravity waves is a long-standing problem of significant theoretical and applied interest. Recently, \citet[\emph{J. Fluid Mech.}, vol. 841,][pp. 463-488]{barthelemy2018on-a-unified-br} examined the energetics of focusing wave groups in deep and intermediate depth water and found that breaking and non-breaking regimes were clearly separated by the normalised energy flux, $B$, near the crest tip. Furthermore, the transition of $B$ through a generic breaking threshold value $B_\mathrm{th} \approx 0.85$ was found to precede visible breaking onset by up to one fifth of a wave period. This remarkable generic  threshold for breaking inception has since been validated numerically for 2D and 3D domains and for shallow and shoaling water waves; however, there is presently no theoretical explanation for its efficacy as a predictor for breaking. This study investigates the correspondence between the parameter $B$ and the crest energy growth rate following the evolving crest for breaking and non-breaking waves in a numerical wave tank using a range of wave packet configurations. Our results indicate that the time rate of change of the $B$ is strongly correlated with the energy density convergence rate at the evolving wave crest. These findings further advance present understanding of the elusive process of wave breaking.
\end{abstract}

%\section*{Key points}
%\begin{itemize}
%\item A numerical study is used to investigate the links between the breaking onset parameter $\mathbf{B}$ and the energy density $E$
%\item The evolution of $E$ is demonstrably different for non-breaking and breaking waves
%\item The rate of change of $\mathbf{B}$ at breaking inception ($\mathbf{B}=B_\mathrm{th}$) is strongly correlated to the rate of change of $E$. This rate of change represents the rate of local energy density convergence and is denoted $\Gamma_E$ 
%\end{itemize}

\section{Introduction}
The physical process of wave breaking remains one of the classical unresolved problems of fluid dynamics. Considerable research effort has been devoted to this topic, but the highly nonlinear nature of the breaking process makes both observational and numerical efforts challenging. Over the years a number of diagnostic parameters have been proposed to characterise the breaking onset process, with \citet{perlin2013breaking-waves-} providing the most recent review of progress in this field. While many parameters have been successful in characterising breaking onset for a specific subset of surface gravity waves, until recently none have proven to apply generically across a range of water depths, generation or instability mechanisms. 

An approach that has shown considerable promise is based upon the evolution of the intragroup energy flux \citep{tulin2001breaking}. The key physical concept is that breaking onset in an unsteady wave group is triggered when the energy flux convergence rate, as measured in a frame co-moving with the tallest crest in the group, exceeds a local stability level \citep{banner_peirson_2007, derakhti2016breaking-onset-}. Recently, \citet[hereafter B18]{barthelemy2018on-a-unified-br} proposed a breaking inception parameter $B$ that links the local energy flux
\begin{equation}
    \mathbf{F} = \mathbf{u}\left(\left(p-p_0\right) + \rho g z + 1/2\rho \lVert\mathbf{u}\rVert^2 + E_0\right)
\end{equation}
to the local energy density
\begin{equation}
   E = \rho g z + 1/2\rho\lVert\mathbf{u}\rVert^2 + E_0.
\end{equation}
Here $p$ and $p_0$ are the pressure within the fluid and at the interface, $z$ is the vertical coordinate, $\rho$ is the density, $g$ the gravitational acceleration and $\mathbf{u}$ the fluid velocity. As  $z=0$ at the still water level, a reference energy term $E_0 = -\rho g z_0$ ensures that $E$ is always positive. The constant $z_0$ is set to twice the water depth, the choice of which has been previously shown to have negligible impact on results \citepalias{barthelemy2018on-a-unified-br}. 

The normalised local energy flux within the crest is derived by dividing these quantities by the wave crest speed $\mathbf{c}$,
\begin{equation}\label{eq:B}
    \mathbf{B} = \frac{\mathbf{F}}{E\lVert\mathbf{c}\rVert}
\end{equation}
and the breaking inception parameter $B$ is defined as $B=\lVert\mathbf{B}\rVert$. \citetalias{barthelemy2018on-a-unified-br} found that, when tracking $B$ for any particular crest, the transition of $B$ through the generic breaking inception threshold level of $B_\mathrm{th}\approx 0.85$ separates breaking and non-breaking wave crests. For those crests that exceed the threshold $B_\mathrm{th}$, breaking onset -- defined as the instant that the crest interface height becomes multi-valued -- is observed when $B > 1.0$. Thus the breaking inception threshold of $B = B_\mathrm{th}$ represents a point of no return beyond which breaking will occur. This picture of wave breaking has been subsequently validated in laboratory studies \citep{saket2017on-the-threshol,SAKET2018159}, with a variety of wave packet types \citep{DerakhtiMorteza2018Ptbs}, for deep water waves \citep{seiffert2018simulation-of-b} and for waves shoaling on topography \citep{Derakhti_2020}. 

As a diagnostic parameter, $B$ has practical advantages. At the water surface $p-p_0=0$ and the breaking inception parameter reduces to $B=\lVert\mathbf{u}\rVert/\lVert\mathbf{c}\rVert$, which can be measured in a laboratory or field setting. In addition, the horizontal components of $\mathbf{B}$ and $\mathbf{c}$ are much larger than the vertical components such that $B_x=F_x/(Ec_x)$ can be used in place of $B$ to a close approximation. $B$ has also demonstrated potential as a forecasting parameter, with the breaking inception threshold $B_\mathrm{th}$ exceeded up to half a wave period prior to breaking onset \citepalias{barthelemy2018on-a-unified-br}. 

While the formulation of $B$ is a dynamic threshold, the underlying reason that breaking occurs only for wave crests in which $B$ exceeds $B_\mathrm{th}$ is yet to be determined. To advance towards resolving this knowledge gap, we consider the main component of (\ref{eq:B}), the local mechanical energy density $E$, in isolation. We investigate the hypothesis that the breaking inception threshold  mimics the local energetics at the crest tip. We track the evolution of both $B$ and $E$ as the crest either relaxes from its maximum steepness without breaking, or transitions to breaking onset. Our experiment utilises direct numerical simulation with a two-phase volume-of-fluid Navier-Stokes solver \citep{popinet2003gerris:-a-tree-,popinet2009an-accurate-ada} to examine fully nonlinear wave packets in the presence of viscosity and surface tension. Using these results we study the temporal evolution of the energy density and compare this to the evolution of $B$. 

\section{Theoretical background}

The determination of breaking inception using $B$ is achieved by tracking the maximum value of $B$ within the crest, which occurs at or near the crest tip. As the crest is propagating at velocity $\mathbf{c}$ the temporal evolution of $B$  has both a local component and a component in the frame of reference of the moving crest \citep[equation (1.2)]{tulin-2007},
\begin{equation}\label{eq:DBDt}
    \frac{D_{c}B}{Dt} 
    = \frac{\partial B}{\partial t} + \mathbf{c} \cdot \nabla B 
    = - \nabla \cdot \left( \left[\mathbf{u} - \mathbf{c}\right]B\right),
\end{equation}
where $D_c / Dt$ denotes the rate of change in the unsteady crest-following frame of reference. 

\cite{DerakhtiMorteza2018Ptbs} found that the strength of breaking is proportional to the rate of change of $B$ at breaking inception ($B = B_{\mathrm{th}}$). They defined the parameter
\begin{equation}\label{eq: GammaB_def}
\Gamma_B = T_0 \left. \frac{D_c B}{Dt}\right|_{B_\mathrm{th}}
\end{equation}
where the rate of change of $B$ is normalised by the local crest period $T_0$. (Note that $B$ and $\Gamma_B$ are both dimensionless quantities.) (\ref{eq: GammaB_def}) is a significant finding as it shows that not only does $B$ provide advance warning of breaking, it is also indicative of the strength of the breaking and the energy dissipation thereafter. 

In the same manner as (\ref{eq:DBDt}), in the crest-following frame, the local energy balance \citep[equation (2.3.2)]{phil1977} can be expressed as
\begin{equation}\label{eq:energy-balance}
    \frac{D_{c}E}{Dt} 
    = \frac{\partial E}{\partial t} + \mathbf{c \cdot \nabla} E 
    = \mathbf{u \cdot f} - \mathbf{\nabla \cdot F_c},
\end{equation}where $\mathbf{u \cdot f}$ is a sink term representing the work done against friction and $\mathbf{F_c}=\left(\mathbf{u}-\mathbf{c}\right)E + \mathbf{u}\left(p-p_0\right)$ is the divergence of the energy flux in the crest-following frame. 

Our aim is to relate the behaviour of $B$, particularly its rate of change following the crest tip $D_cB/Dt$, to the more familiar wave energy growth rate $D_cE/Dt$ following the crest tip. We introduce the normalised growth rate:
\begin{equation}\label{eq:Gamma-E}
\Gamma_E=\frac{T_0}{E - E_0} \left.\frac{D_cE}{Dt}\right|_{B_{\mathrm{th}}}.
\end{equation}
In (\ref{eq:Gamma-E}), the local rate of change of $E$ following the crest tip is normalised by the dynamic local energy density ($E - E_0$), divided by the local crest period $T_0$. The arbitrary reference energy level $E_0$, which does not affect $D_cE/Dt$, is suppressed in the denominator to allow a generic comparison of deep and shallow water cases. 

\section{Experiment description}

To elucidate the relationship between $\Gamma_E$ and $\Gamma_B$, we conducted a suite of numerical simulations of breaking and non-breaking waves across a range of wave packet configurations and grid refinements (table \ref{tab:experiments}). We used the Gerris software package \citep{popinet2003gerris:-a-tree-} to numerically solve the two-dimensional,  incompressible, variable density Navier-Stokes equations, including the effects of viscosity and surface tension. Gerris uses the Volume-Of-Fluid (VOF) method to simulate two-phase flows, with surface tension modelled through an improved implementation of the continuum-surface-force approach \citep{popinet2009an-accurate-ada}. Gerris has been extensively validated for simulations of surface gravity waves \citep{WRONISZEWSKI20141}, wave breaking kinematics \citep{deike2017lagrangian-tran,pizzo2016current-generat} and energy dissipation \citep{de2018breaking}.

\begin{figure}
  \centerline{\includegraphics[width=\textwidth]{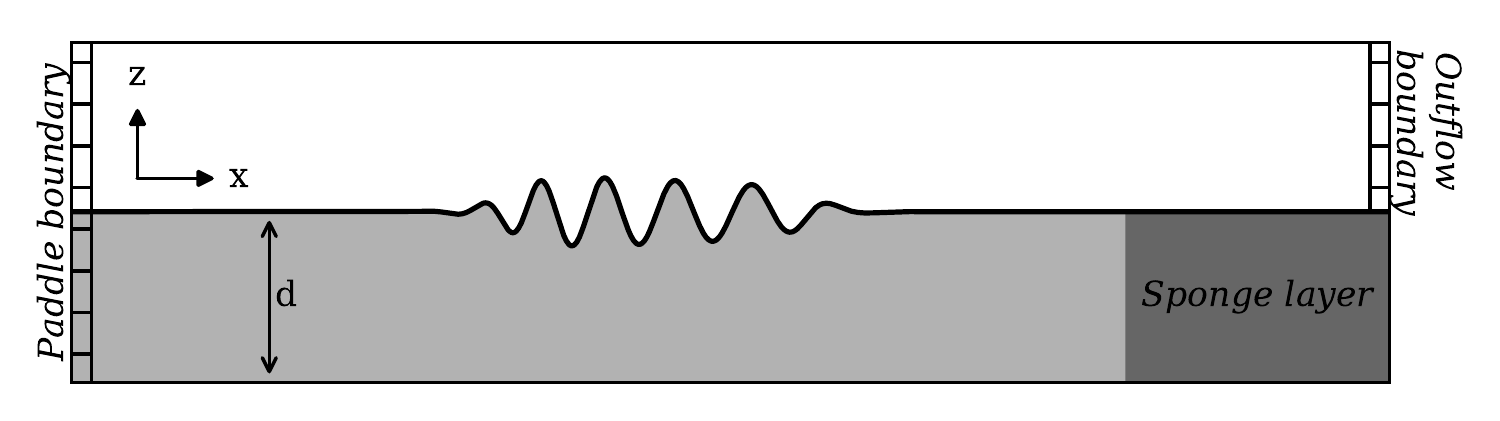}}
  \caption{The numerical wave tank. Waves are generated at the paddle boundary and travel down the tank in the positive $x$ direction. The depth of water $d$ is varied to achieve the desired depth/wavelength ratio. A numerical sponge layer absorbs waves at the far end of the tank, while an outflow boundary minimises pressure gradients in the air phase due to paddle movement. A typical chirped wave packet ($N=5$, enlarged for clarity) is shown.}
\label{fig:tank}
\end{figure}

The model is set up as a two-dimensional numerical wave tank of length $23.5\lambda_p$ and height $1.18\lambda_p$, where $\lambda_p$ is the deep-water wavelength of the wave paddle forcing (figure \ref{fig:tank}). While computational constraints limit us to two-dimensional simulations, previous studies have shown that there is negligible difference in  $B$ between two- and three-dimensional cases \citep{barthelemy2018on-a-unified-br,DerakhtiMorteza2018Ptbs}.

Waves are generated at the left-hand side of the tank. We simulate a bottom-mounted flexible flap paddle by deriving the exact solutions for velocity and pressure gradient from wavemaker theory \citep{dean1991water-wave-mech} and apply these at the fixed boundary. This method greatly increases the computational efficiency of the model while still generating a fully nonlinear wave packet. The lateral movement $A_p$ of the simulated paddle is $<5 \% $ of the wavelength in most cases (table \ref{tab:experiments}) so the approximation of a fixed boundary has little effect on the results. 

The motion of the paddle $x_p$ with time $t$ follows the chirped packet function \citep{song2002on-determining-},
\begin{multline}
x_p(t)=-0.25A_p\left(1+\tanh\left[\frac{4\omega_pt}{N\pi}\right]\right)\left(1-\tanh\left[\frac{4\left(\omega_pt-2N\pi\right)}{N\pi}\right]\right)\\ 
\times\sin\left(\omega_p t\left[1-\frac{\omega_pC_{ch}t}{2}\right]\right)\, 
\end{multline}
where $x_p$ is a function of the paddle forcing amplitude $A_p$, the forcing frequency $\omega_p$, the number of waves in the packet $N$ and the packet linear chirp rate $C_{ch}=1.0112 \times 10^{-2}$. We vary $A_p$ and $N$ to generate an ensemble of non-breaking and breaking waves (table \ref{tab:experiments}) of varying amplitude and breaking strength. 

Energy absorption at the far end of the tank is achieved through a number of complementary approaches. The final $4.7\lambda_p$ of the tank consists of a numerical sponge layer based on that derived by \cite{clement1996coupling-of-two}, which effectively absorbs high frequency waves. The reflection of low frequency waves is minimised by gradually increasing the grid spacing within the sponge layer to enhance numerical dissipation. An outflow boundary condition is also applied to the dry portion of the lateral boundary to minimise compression of the air phase caused by the paddle motion, which further improves the performance of the model's Poisson solver. 

Gerris uses a quadtree mesh structure which enables efficient adaptive mesh refinement \citep{popinet2003gerris:-a-tree-}. Each level of refinement divides the parent cell into four, resulting in a maximum resolution equivalent to an uniform mesh size of of $2^j \times 2^j$, for $j$ refinement levels. As our primary interest in this study is focused on the air-water interface and the water boundary layer, we determine the maximum required resolution based on the boundary layer thickness $\delta={\lambda_p}/{\sqrt{Re}}$ \citep{phil1977} where $Re=\rho U\lambda_p / \sigma$ is the wave Reynolds number. To reduce computational cost we set $Re = 4 \times 10^4$ which allows us to resolve the boundary layer with four cells at a refinement level of $2^{10}$ and equates to a resolution of $\lambda_p/870$ with the scaling used. While this is smaller than our physical $Re = 1.25 \times 10^6$, previous studies \citep{deike2017lagrangian-tran, mostert_deike_2020} have shown that $Re = 4 \times 10^4$ is large enough that viscous effects are not dominant and all energy within the boundary layer is adequately resolved. A limited number of experiments with a maximum refinement level of $2^{11}$ (eight cells within the boundary layer, equivalent to $\lambda_p/1750$) are also reported on in the following section. For all experiments, mesh refinement criteria are configured to ensure maximum resolution at the air-water interface and in regions of large vorticity.

A total of 74 experiments were completed with a range of resolution, wave packet size, water depth, and paddle amplitudes (table \ref{tab:experiments}) generating an ensemble of 285 non-breaking and 52 breaking crests for analysis. All parameters are presented as non-dimensional quantities.
\begin{table}
  \begin{center}
\def~{\hphantom{0}}
  \begin{tabular}{ c c c c c }
    Refinement level    & $N$   & $d/\lambda_p$	& No. of cases  & $A_p/\lambda_p$  \\[3pt]
    $2^{10}$     	    & 5     & 0.59          & 27   		    & $0.0250-0.0500$\\
    $2^{10}$            & 9     & 0.59          & 32  	        & $0.0250-0.0450$\\
    $2^{10}$            & 5     & 0.20          & 9            & $0.0800 - 0.0920 $\\
    $2^{11}$            & 5     & 0.59          & 3		        & $0.0370-0.0460$ \\
    $2^{11}$            & 9     & 0.59          & 3		        & $0.0370-0.0389$  
  \end{tabular}
  \caption{Summary of experiments included in this study. The model was configured using a range of mesh refinement levels, wave packets $N$ and water depth $d/\lambda_p$. For each configuration the amplitude of the paddle $A_p/\lambda_p$ was varied to generate an ensemble of breaking and non-breaking crests.}
  \label{tab:experiments} 
  \end{center}
\end{table}

\section{Results}
\begin{figure}
  \centerline{\includegraphics[width=\textwidth]{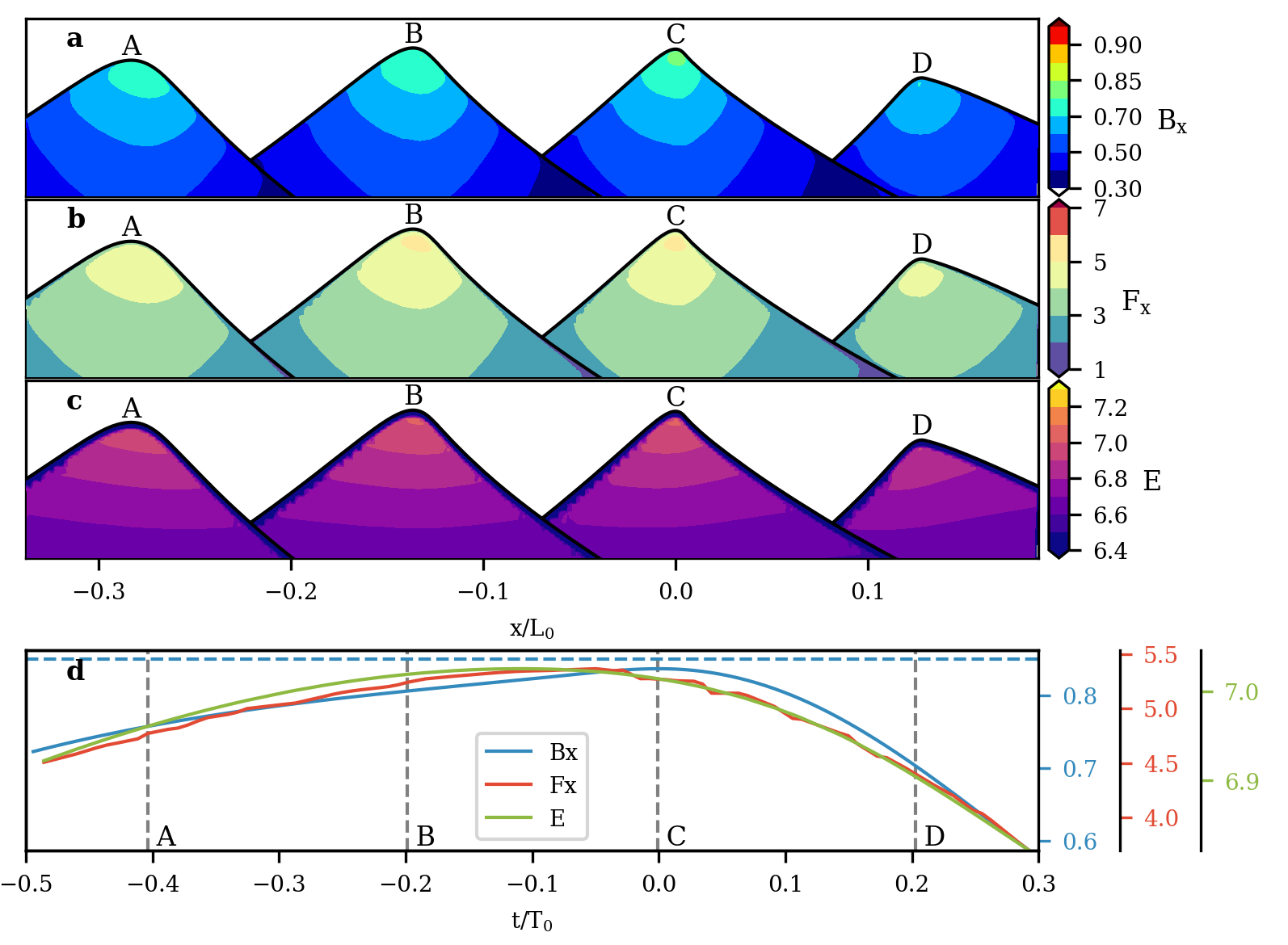}}
  \caption{The evolution of $B_x$ (a), $F_x$ (b) and $E$ (c) for a non-breaking wave progressing through the growing and decaying phase. The horizontal axis is normalised by the deep-water wave length $L_0$ and the vertical axis is exaggerated by a factor of $10:1$. The temporal evolution of the maximum crest value of each parameter is shown in panel d. The time of each snapshot (A-D) is indicated by the dashed lines. Times are normalised by the instantaneous deep-water crest period $T_0$ and referenced to the time of maximum $B$. The value of $B_\mathrm{th}$ is also indicated by the blue dashed line in panel d.}
\label{fig:nb-slice}
\end{figure}

We first examine the evolution of the critical parameters $B_x$, $F_x$ and $E$ for a maximally recurrent non-breaking wave (figure \ref{fig:nb-slice}). Snapshots of the wave evolution before, at, and after the time of maximum $B$ are shown. For each parameter, a local maximum is visible at the crest of the wave. In studies utilising an inviscid solver \citep{SeiffertBr2017Sobw, barthelemy2018on-a-unified-br} these maxima are located at the crest surface. In our simulations, where the impacts of viscosity and surface tension are included, we find that the maxima occur at the edge of the interfacial boundary layer. A consequence of limiting the Reynolds number to $4 \times 10^4$ and effectively increasing the thickness of the turbulent boundary layer is that the depth of the maxima below the interface is amplified. However, in other aspects, such as the magnitude of $B$, our results are consistent with those previous studies.

At each time, the position and magnitude of the maxima is located with a two-dimensional spline to derive the temporal evolution of these crest values (figure \ref{fig:nb-slice}d). Times are normalised by the local crest period $T_0$ and referenced to the time of maximum $B$ (which we set to be $t = 0$). While the absolute values of $F_x$ and $E$ differ, their evolution in time are very similar. As would be expected, the evolution of $B_x$ is closely related; however, the time of the peak value occurs slightly later than $F_x$ and $E$ due to the dependence on the crest speed $c$ (equation (\ref{eq:B})), which undergoes a regime of deceleration and acceleration as the crest evolves \citep{banner2014linking-reduced, fedele_banner_barthelemy_2020}.

The crest speed $c$ is a critical parameter in the calculation of $B$ but it is difficult to calculate accurately \citep{Derakhti_2020}. We achieve this by firstly applying a smoothing filter to the interface, which removes small-scale ripples. A low-pass filter is then applied to the resultant crest positions, and a smooth cubic spline used to interpolate between data points. Comparison of the smoothed crest position with the evolution of the interface confirms that this is a robust method for calculating the crest speed.  

\begin{figure}
  \centerline{\includegraphics[width=\textwidth]{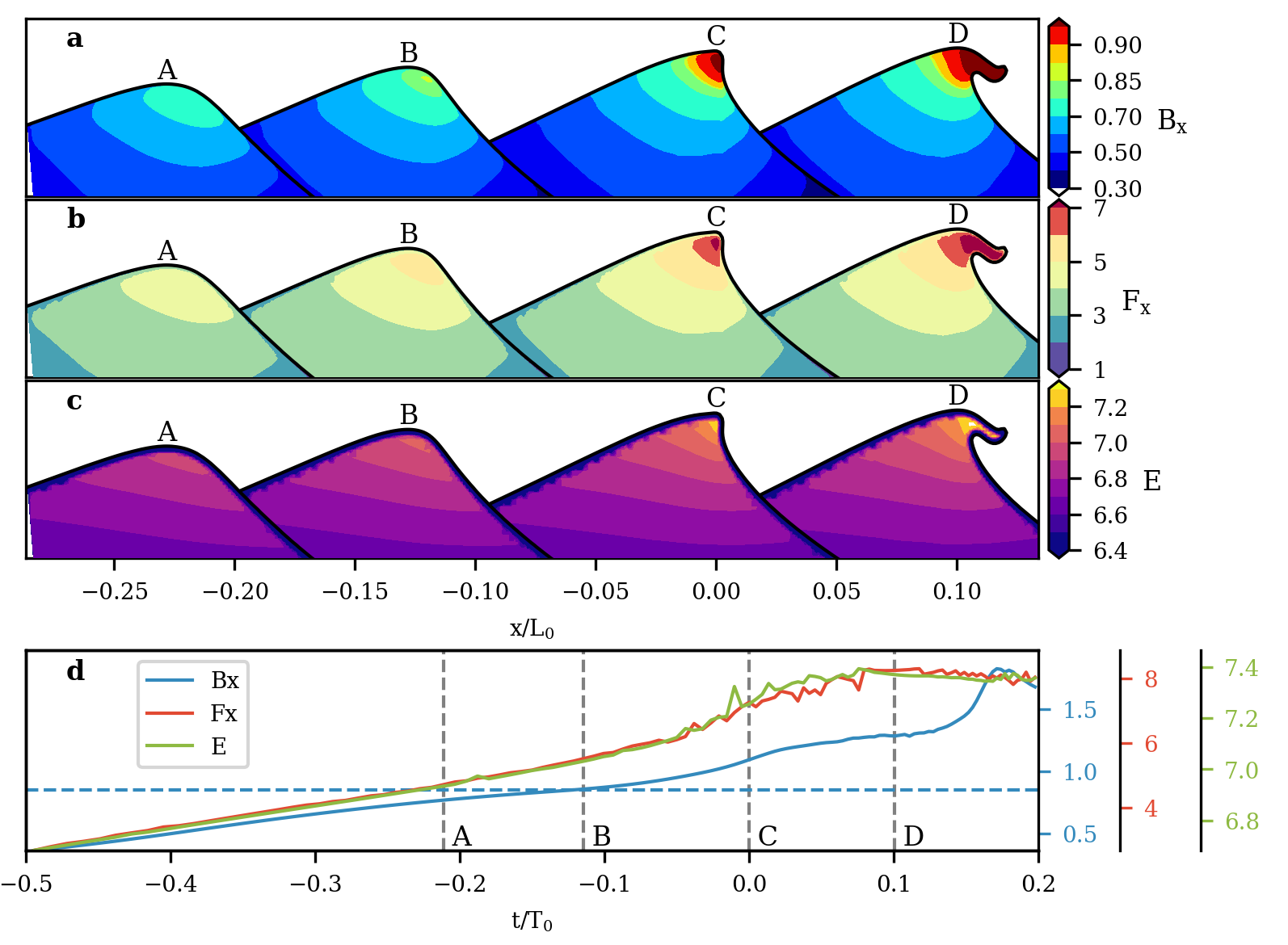}}
  \caption{As for figure \ref{fig:nb-slice} but for a breaking wave. The vertical axis is exaggerated by a factor of $8:1$. Times are referenced to the time that breaking is first detected.}
\label{fig:b-slice}
\end{figure}

In the breaking case (figure \ref{fig:b-slice}) the local maxima of each parameter are more clearly defined and are located on the forward crest face at the instant of breaking. Breaking onset, defined as the time when the interface height first becomes multi-valued, occurs approximately $0.1 - 0.2$ deep-water wave periods after breaking inception ($B=B_\mathrm{th}$). The rates of change of both $B$ and $E$ at breaking inception  (i.e. $\Gamma_B$ and $\Gamma_E$) are smoothly varying and approximately linear (figure \ref{fig:b-slice}d).  

\begin{figure}
  \centerline{\includegraphics[width=\textwidth]{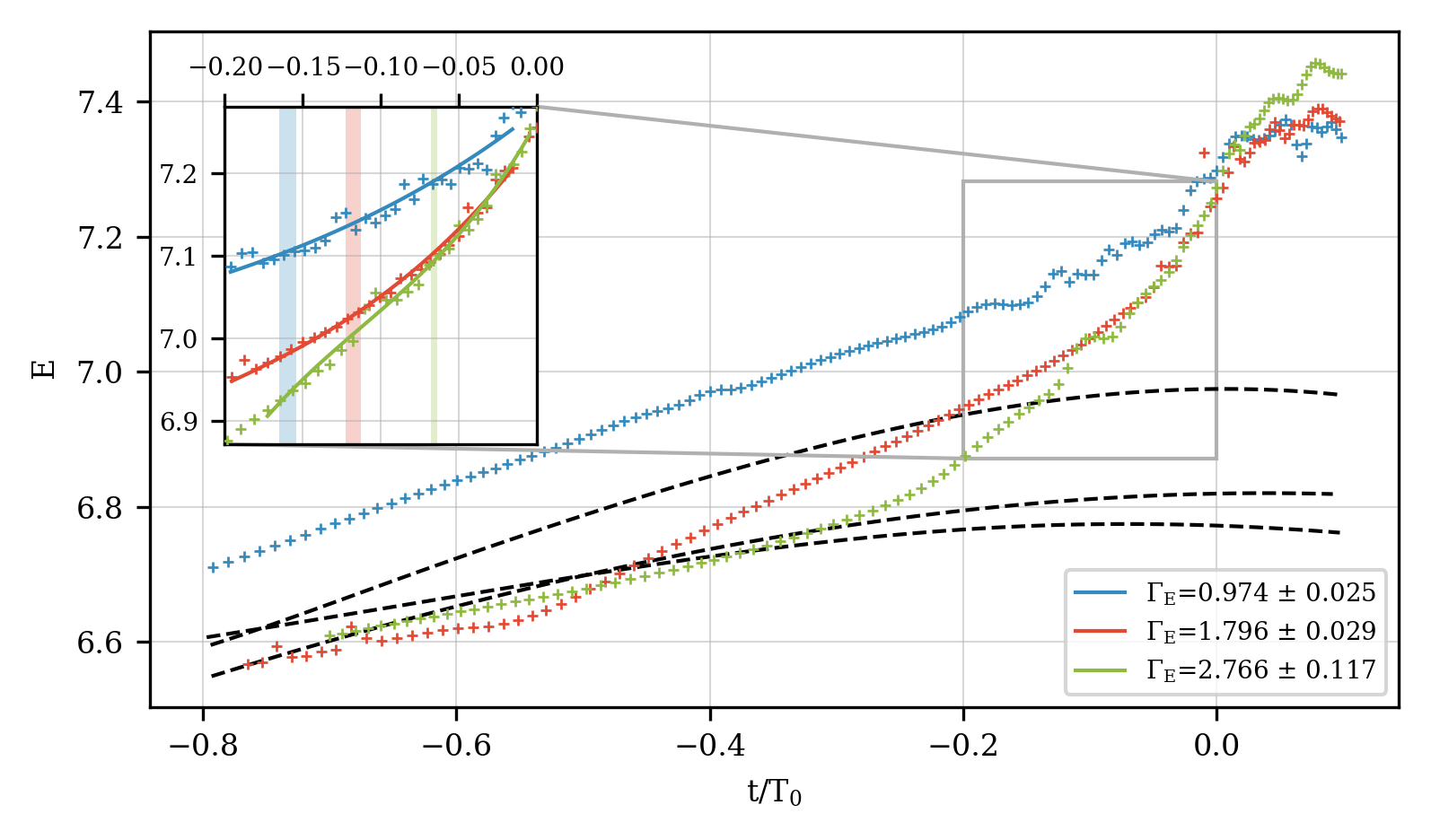}}
  \caption{Evolution of the local energy density $E$ for non-breaking (black), weak (blue), moderate (red) and strong (green) breaking crests. Time is relative to the maximum $B_x$ value (non-breaking crests), or to the time of breaking onset (breaking crests). The calculation of $\Gamma_E$ (inset) is done by fitting a smooth spline over the time period that $0.75 < B < 1.0$. $\Gamma_E$ is taken as the average slope of the spline for the time period that $0.85 < B < 0.86$ (coloured shaded region) and is reported to $95\%$ confidence.}
\label{fig:E-time-series}
\end{figure}

We examine the evolution of $E$ in more detail in figure \ref{fig:E-time-series}. For non-breaking waves, the magnitude of $E$ plateaus as the maximum $B$ value is reached ($t/T_0=0$), with the peak value of $E$ increasing as a function of the maximum wave amplitude. Conversely, in a breaking crest $E$ continues to increase through breaking inception and past breaking onset. While the absolute range of $E$ is small, there is a distinct separation in values at $t/T_0 =0$ between the non-breaking and breaking crests.

For breaking waves the energy density convergence rate $\Gamma_E$ (equation (\ref{eq:Gamma-E})) is calculated by first fitting a local smooth spline to the $E$ time-series (figure \ref{fig:E-time-series} inset). The spline is fit over the time interval for which $0.7 < B < 1.0$, chosen to optimise the spline fit for the period of interest while also capturing any variability in $E$.  The first derivative of the spline yields $D_cE / Dt$ (equation (\ref{eq:energy-balance})); $\Gamma_E$ is then taken as the normalised value of $D_cE / Dt$ as the crest passes through $B_\mathrm{th}$. To account for the uncertainty in the absolute value of $B_\mathrm{th}$, $D_cE / Dt$ is averaged over the interval  $0.85 < B < 0.86$ (shaded regions in figure \ref{fig:E-time-series} inset). $D_cE / Dt$ is nearly constant at this time and we find that $\Gamma_E$ is relatively insensitive to the choice of averaging interval. 

The three breaking examples shown in figure \ref{fig:E-time-series} are characterised as weak, moderate and strong breaking crests based on the magnitude of $\Gamma_B$ (here calculated using an equivalent method to $\Gamma_E$). It can be seen that the magnitude of $\Gamma_E$ correspondingly increases with increasing $\Gamma_B$. This is an interesting result as nothing else appears to distinguish the evolution of $E$ between these cases; there is no trend in the value of $E$ at breaking inception and $E$ is nearly identical in all three cases at breaking onset. 

\begin{figure}
  \centerline{\includegraphics[width=1\textwidth]{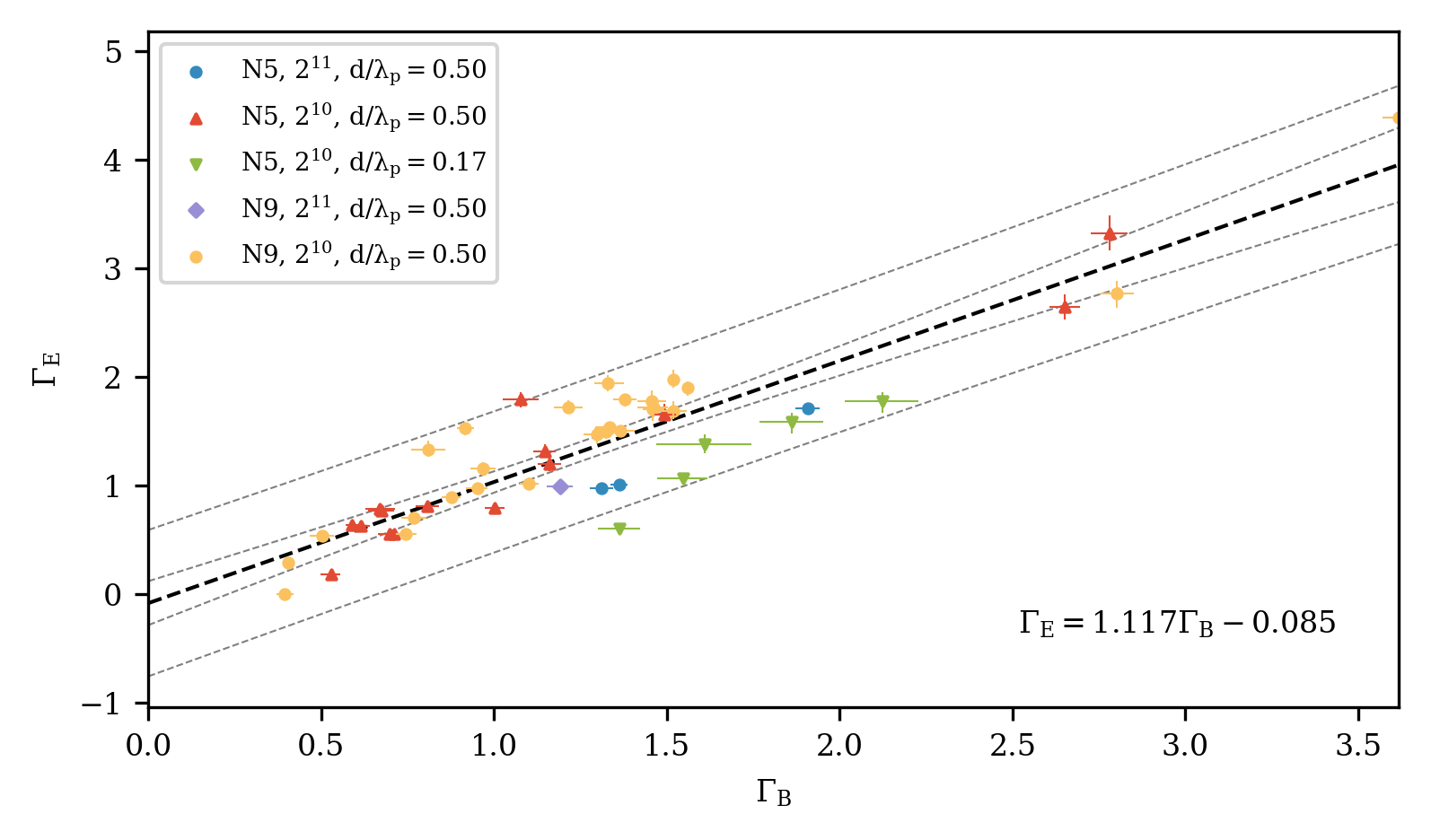}}
  \caption{Relationship between $\Gamma_B$ and $\Gamma_E$  for all breaking crests from the experiments listed in table \ref{tab:experiments}. Error bars indicate the sensitivity to varying the averaging interval when calculating $\Gamma$. A linear regression has been applied (black dashed line) with the grey dashed lines indicating the $95\%$ confidence (inner) and prediction (outer) intervals.}
\label{fig:Gamma-comparison}
\end{figure}

The strong link between $\Gamma_E$ and $\Gamma_B$ is seen across all breaking crests in our ensemble, regardless of wave packet configuration, water depth or model resolution (figure \ref{fig:Gamma-comparison}). The robustness of the relationship was further tested by varying the averaging period used in the calculation of $\Gamma$ between $0.84 < B < 0.85$ and $0.86 < B < 0.87$, with no significant impact on the results. 

\section{Discussion and conclusions}
The aim of this study has been to make progress towards a physical explanation as to why the breaking inception parameter $B$ is a reliable predictor of breaking. The threshold of $B_\mathrm{th}\approx 0.85$ separating breaking and non-breaking waves first reported by \citet{barthelemy2018on-a-unified-br}  has since been confirmed in further independent studies. A significant feature of $B_\mathrm{th}$ is that it provides advanced warning of breaking onset --- up to $0.2$ deep-water wave periods in our results.  \cite{DerakhtiMorteza2018Ptbs} shed further light on the subject by showing that the normalised rate of change of $B$ at breaking inception, $\Gamma_B$, is strongly correlated to the strength of the eventual breaking event. 

We have used direct numerical simulation to investigate the links between $B$ and the local crest energy density $E$. In an ensemble of experiments spanning a range of wave packets, water depths and model resolutions, we have shown that the crest energy growth rate, $\Gamma_E$, is strongly correlated to $\Gamma_B$ and is therefore also an indicator of the breaking strength. 

We now move to a discussion on the physical interpretation of these results. Equation (\ref{eq:energy-balance}) links the divergence of the energy flux to the rate of change of the energy density. As the work done against friction is small compared to the energy flux divergence (here $\lVert \mathbf{u} \cdot \mathbf{f} \rVert / \lVert \nabla \cdot \mathbf{F_c} \rVert  =O\left(10^{-3}\right)$) then (\ref{eq:Gamma-E}) can also be expressed as
\begin{equation}\label{eq:Gamma-E-as-div}
\Gamma_E\approx - \frac{T_0}{E - E_0} \left.\nabla \cdot \mathbf{F_c}\right|_{B_\mathrm{th}}.
\end{equation}
Thus, $\Gamma_E$ represents the energy flux convergence within the crest and is closely related to the mechanism that leads to breaking: an excessive flow of energy into the crest triggers a local instability which can only be dissipated through the process of breaking. However, while $\Gamma_E$ provides a physical explanation for the process of breaking inception, the highly nonlinear nature of the breaking process makes $\Gamma_E$ difficult to quantify except via a detailed numerical simulation. 

In this study, we have re-examined the energetics of wave breaking onset through the lens of the breaking inception parameter, $B$, which is related to the normalised energy flux near the crest tip. We have shown that $\Gamma_B$ (the rate of change of $B$) is an effective proxy for the energy growth rate, $\Gamma_E$. Since $B = \lVert \mathbf{u} \rVert / \lVert \mathbf{c} \rVert$ at the crest surface, both $B$ and $\Gamma_B$ can be readily measured in a laboratory or field experiment. We therefore see that the utility of the inception parameter $B$ as a predictor of wave breaking derives from its close relation to the energy flux convergence near the wave crest, which is the underlying physical process leading to breaking onset. However, an explanation for the existence of the generic breaking inception threshold $B_\mathrm{th}\approx 0.85$ remains to be determined.

\section*{Acknowledgements}
This research was supported by resource grants under the National Computational Merit Allocation Scheme (NCMAS) and the Intersect Compute Merit Allocation Scheme (ICMAS). DB is supported by an Australian Government Research Training Program (RTP) Scholarship.

\section*{Declaration of Interests} The authors report no conflict of interest.
\bibliographystyle{jfm}
\bibliography{reference-library}

\end{document}